\documentclass{eptcs}
\usepackage[utf8x]{inputenc}
\newif\ifdraft
\draftfalse

\usepackage{hyperref}
\usepackage{cleveref}
\hypersetup{pdfborder ={0 0 0}, colorlinks=false}
\usepackage{listings}
\usepackage{xparse}
\usepackage[english]{babel}
\usepackage{amssymb}

\usepackage{xspace}
\usepackage{empheq}
\usepackage{stmaryrd}                                                                                                         \usepackage{multicol}
\usepackage{mathtools}      
\usepackage[vlined,algoruled,linesnumbered]{algorithm2e} 
\usepackage[table]{xcolor}
\usepackage{longtable} 
\usepackage{booktabs}      

\usepackage{graphicx} 
\usepackage{comment}

\usepackage{eurosym}

\usepackage{subfig}
\usepackage{wrapfig}
\usepackage{framed}
\usepackage[strict]{changepage}

\usepackage{mdwlist}
\usepackage{enumitem}

\DeclareMathOperator*{\opcoeur}{\heartsuit}
\NewDocumentCommand{\coeur}{O{}m}{\opcoeur_{#1}\left(#2\right)}

\newcommand*{\lab}[1]{\ensuremath{\coeur{#1}}}
\DeclareMathOperator*{\opRHSs}{R}
\newcommand*{\RHSset}[2]{\ensuremath{\opRHSs_{#1}\left(\ifthenelse{\equal{#2}{}}{\card{#1}}{#2}\right)}}
\newcommand*{\labelset}[2]{\ensuremath{\lab{#1}\ifthenelse{\equal{#2}{}}{}{\left[:#2-1\right]}}}

\def\Sub#1{\ensuremath{\text{\rm Sub}(#1)}}

\NewDocumentCommand{\hsend}{om}{\ensuremath{\rcv{}{#2}}}
\NewDocumentCommand{\anyrule}{o}{\ensuremath{\mathrel{\rlap{\hspace{.45em}$*$}\rightarrow}}}
\NewDocumentCommand\stdrule{o}{\ensuremath{\rightarrow}}
\NewDocumentCommand\hnoncereceive{om}{\ensuremath{\stdrule #2}}

\NewDocumentCommand{\pscss}{O{T}}{\ensuremath{\Sub{#1}}}
\NewDocumentCommand{\byDec}{O{}O{}}{\ensuremath{\operatorname{byDec}^{#1}_{#2}}}
\NewDocumentCommand{\ruleDec}{O{}O{}m}{\ensuremath{\operatorname{\rho}^{#1}_{#2}\left(#3\right)}}

\NewDocumentCommand{\cmnt}{O{T}m}{\marginpar{\footnotesize  #1: #2}}

\title{Compiling symbolic attacks to protocol implementation tests \thanks{This work is  supported by  FP7 NESSoS 
 project and DAST Investissement d'Avenir project.}
}
 \author{Hatem Ghabri~\institute{Sup Com, Tunis} \email{hatem.ghabry@gmail.com} \and   Ghazi Maatoug \institute{Sup Com, Tunis}  
\email{ghazi.maatoug88@gmail.com} \and   Micha{\"e}l Rusinowitch 
  \institute{INRIA Nancy Grand Est} \email{rusi@inria.fr}
   }

\def\titlerunning{Compiling Symbolic Attacks}
\def\authorrunning{H. Ghabri, G. Maatoug \& M. Rusinowitch}
\Crefname{proposition}{Proposition}{Propositions}
\crefname{proposition}{proposition}{propositions}

\Crefname{corollary}{Corollary}{Corollaries}
\crefname{corollary}{corollary}{corollaries}

\Crefname{definition}{Definition}{Definitions}
\crefname{definition}{definition}{definitions}

\Crefname{algocf}{Algorithm}{Algorithms}
\crefname{algocf}{algorithm}{algorithms}

\excludecomment{removable}

\begin{document}

\maketitle
\begin{abstract}
Recently efficient model-checking tools have been developed to find flaws in security protocols  specifications. 
These flaws  can be interpreted as potential attacks scenarios  
but the feasability of these scenarios  need to be confirmed at the implementation level. 
However, bridging the gap between an abstract attack scenario derived from a specification  and a penetration test 
on real implementations of a protocol is still an open issue.
This work investigates an architecture for  automatically generating abstract attacks  and converting them 
to concrete tests on protocol implementations. 
In particular we aim to improve  previously proposed blackbox testing methods in order to discover automatically 
new attacks and vulnerabilities.  As a proof of concept
we have experimented our proposed architecture to detect a renegotiation vulnerability  on some 
implementations of SSL/TLS,  a  protocol widely used for securing electronic transactions.  
\end{abstract}


\section{Introduction}

Personal, commercial and administrative  communications are   increasingly 
routed  through open networks like Internet and rely on security  protocols such as SSL 
to protect their privacy. However new vulnerabilities are discovered  every day  by hackers 
on these protocols based on implementation flaws or improper  parameter setting.

Formal verification of cryptographic protocols \cite{rewriting,cervesato,chevalier01,avispa,Gawanmeh} is a widely developed area but it is insufficient in general 
to ensure  that the protocol  implementations are secure.  
Technical details are often abstracted in formal verification and they are the source of many 
vulnerabilities. 
Moreover implementations 
often  diverge from the initial specification for instance when the implementer misinterprets ambiguous documentations. 

Penetration testing 
permits the test of implementations based on the tester initial knowledge. 
The tester tries to get unto the system by performing manually attack scenarios. Depending, on its 
knowledge  of the protocol implementation  we find three major category: White Box, Grey Box and Black Box testing. 
Different from  penetration testing, fuzzing techniques bombs the protocol implementation 
with random inputs which are not conform with the protocol specification. It is a well known method for finding software failures because it targets the error handling part. 
We need more effective methods to improve the efficiency of implementation testing tools, as most of the existing approaches 
resort to random or manual t-esting. Since  model-checkers have been remarkably efficient in 
finding flaws in security protocol specifications  we propose to exploit the power of  these  
formal verification tools for implementation testing too.

\paragraph{Contribution.}
In this work we adopt a model-based approach to protocol testing. 
We start from a secure formal model of the protocol under test. The first phase 
which is not an original  contribution from ours,  follows the    
proposal in  \cite{dadeau} 
to generate plausible flaws in the model by relying for instance  on  well suited mutation techniques  for focusing  on specific 
meaningful security properties. We assume that an executable protocol mutant potentially  witnesses an implementation flaw. 
Then  we use CL-AtSe  model-checker \cite{clatse}  to generate  attack traces that exploit  
specific security vulnerability w.r.t. authentication or confidentiality  on the protocol mutant.
These generated attack traces represent attacks at a formal level, and 
we need to confirm the vulnerabilities  at the implementation level. 
However, bridging the gap between an abstract attack trace and a penetration test 
on real implementations of the protocol is still an open issue.
We propose an architecture for  automatically  compiling abstract attack traces 
to concrete executable tests on protocol implementations. 
In particular we contribute to improve  previously proposed blackbox testing methods by 
providing  more computer assistance to discover  
new attacks and vulnerabilities.  As a proof of concept
In our experiments we have found a vulnerability  on XAMPP version 1.7.2/openSSL v0.9.8k :
The server accepts SSL renegotiation requests initiated by the client.
However more recent SSL versions have been protected against this vulnerability.

The  mutation method involves some guessing. However instead
of being fully random (like in fuzzing) the  mutation
is oriented towards guessing "standard programming mistakes".
Hence the combinatorial problem is reduced. Much less
erroneous models (scenarios) need to be tested against the concrete implementation.
 By using our plateform solution, the whole process of scenarios translating to concrete tests is automated, however, it is worth to say that a preliminary work have to be mannually done. It consist in mannual configuration of the test environment which depends on the test scenario. For brevity, this paper mainly concentrates on the description of the proposed approach and on its validation by means of a real-life size case study. Formal definitions, more implementation details and other case studies will be included in the extended next version of this paper.   
 
\paragraph{Related works.}
Recent works have proposed several interesting mutation techniques \cite{dadeau}  
for cryptographic protocol specifications. However these method  are not yet 
harnessed in order to compile towards concrete implementation tests. 
A similar work to ours has been published independently  (but after Hatem Ghabri's master \cite{masterghabri}) proposing 
also to compile abstract attacks to concrete penetration tests on implementation  \cite{armando}. 
They  derive a  message construction procedure to compile 
their attacks but  we show that we can  reuse directly an existing  efficient 
and complete procedure that is implemented in Avantssar platform ~\cite{avantssar}.  
The method in \cite{armando}  has been applied to different protocols from ours too. 


\section{SSL/TLS formal modelling and attack trace generation}

\subsection{Overview of SSL/TLS protocol}
\paragraph{}Secure Sockets Layer (SSL) or its updated version Transport Layer Security (TLS) is a cryptographic protocol designed to provide communication security over the internet.
TLS encrypts all the traffic data at the Application Layer for the Transport Layer, using asymmetric cryptography for key exchange, symmetric encryption for privacy, and message authentication codes integrity. Protocols such as HTTP, SMTP; POP3 and others use TLS to create secure connections.

\paragraph{Handshake Protocol} In order to get an SSL session with a server, the Client, 
a Web browser, proceeds by following a handshake sequence described by the handshake protocol.
During the handshake there is a negotiation of session information between the client and the server. This information consists of a session ID, peer certificates, the cipher specification, the compression algorithm and a shared secret that is used to generate keys. The Cipher Specification indicates what methods to use for Key Exchange, Data transfer and creation of message authentication Code, MAC.




Below we refer to the client as A (Alice) and the server as B (Bob) as it is usual for authentication protocols.
At the start of a handshake, A sends a Client Hello message to B supplying a session identifier Sid and a nonce Na, called client random.  Client Hello message also contains a cryptosuite offer Pa which is set of A\'s preferences for encryption and compression.

\begin{equation}
Client \ hello: A \rightarrow B: A, Na, Sid, Pa
\end{equation}  

In response, the server B sends a Server Hello message, containing his nonce Nb (server random), a cipher suite and a compression method selected from ones proposed by the client. Agent B, also, repeats the session identifier Sid.

\begin{equation}
Server\ Hello: B \rightarrow A: Nb, Sid, Pb
\end{equation}

The server's public key, Kb is delivered in a certificate signed by a trusted third party; it is generally an X.509 certificate.

\begin{equation}
Server\ Certificate: B \rightarrow A: certificate(B,Kb)
\end{equation}

\paragraph{}The client, optionally, sends a certificate message which contains the client's certificate
Then he generates a pre-master-secret PMS, a 48-byte random string, and sends it to B with his public key in a Client Key Exchange.
A optionally sends a certificate verify message to authenticate himself.
\begin{equation}
Client\ Certificate*: A \rightarrow B: certificate (A,Ka)
\end{equation}

\begin{equation}
Client\ Key\ Exchange: A \rightarrow B: \{PMS\}\_ Kb
\end{equation}

\begin{equation}
Certificate\ Verify* : A \rightarrow B: \{Hash\{Nb,B,PMS \} \} \_ inv(Ka)
\end{equation}

\paragraph{}The notation $\{\}_k$  stands for the message X encrypted using the key K.
Now both parties calculate the master-secret M from the nonces and the pre-master-secret using a secure pseudo-random-number function PRF. They calculate session keys and MAC secrets from the nonces and master secret. Each session involves a pair of symmetric keys; A encrypts using one and B encrypts using the other. Similarly, A and B protect message integrity using separate MAC secrets.  Before sending application data, both parties exchange Finished messages to confirm all details of the handshake and to check that clear text parts of messages have not been altered. 
\begin{equation}
Master\ Secret\ M = PRF(PMS.Na.Na) 
\end{equation}

\begin{equation}
Finished Message = Hash\{M.A.B.Na.Pa.Sid\} 
\end{equation}

\begin{equation}
Client\ Finished: A \rightarrow B: \{Finished\}\_ClientK
\end{equation}

\begin{equation}
Server\ Finished : B \rightarrow A: \{Finished\}\_ServerK
\end{equation}

\paragraph{}The symmetric key client is intended for client encryption, 
while server is for server encryption. The corresponding MAC secrets are implicit because our model assumes strong encryption.
\begin{equation}
Clientk = KeyGen(A,Na,Nb,M)
\end{equation}

\begin{equation}
ServerK = KeyGen(B,Na,Nb,M)
\end{equation}

\paragraph{}Once a party has received the other's "Finished" message and compared it with her own,  she is ensured that 
both sides agree on all critical parameters, including M and the preferences Pa and Pb. Now she may begin sending confidential data.
\subsection{Formal model}
\paragraph{} We can model the interactions between two roles  in the SSL/TLS Protocol, 
using the Multiset Rewriting in Dolev-Yao Model:
\begin{tabbing}
$ A \rightarrow B: A, Na, Sid, Pa$\\	       	
$ B \rightarrow A: Nb, Sid, Pb$\\
$ B \rightarrow A: \{ B, Kb\}\_inv(Ks)$\\	
$ A \rightarrow B: \{ A, Ka \}\_inv(Ks)$\\	
$ A \rightarrow B: \{ PMS \}\_Kb$\\
$ A \rightarrow B: \{ H(Nb,B,PMS)\}\_inv(Ka)$\\	
$ A \rightarrow B: \{ Finished \}\_Keygen(A, Na, Nb, M)$\\ 
$ B \rightarrow A: \{ Finished \}\_Keygen(B, Na, Nb, M)$
\end{tabbing}
\paragraph{} The previous notation gives us a clear illustration of the messages exchanged in a normal run of a given protocol. But it is not precise enough to specify execution steps and transitions  of the protocol. 
That is why we  also use 
HLPSL  language \cite{avispa} to specify the actions of each kind of participant as a module. 
For each type of participant, in a protocol, there will be one basic role specifying his sequence of actions. This specification can later be instantiated by one or more agents playing the given role.
Each basic role describes what information the participant can use initially (parameters), its initial state, and ways in which the state can change (transitions). 
The AVISPA platform ~\cite{avispa},  uses back-ends to analyse a model in HLPSL and check if the security goals are satisfied or violated. 
If a security goal of the specification is violated, the back-ends provide a trace which shows if the sequence of events leading up to the violation and displays which goals are violated. The command-line AVISPA Tool outputs attack traces in a textual form we will see later.

\subsection{Mutation techniques}
Mutation  is a technique that consists in introducing faults into a model in order to generate test cases. 
Model checkers can then be employed  to detect flaws caused by the introduced mutation. 
There exist many mutation techniques and we try to present some (see \cite{dadeau}) that are well adapted  to  our context.

\paragraph{Agent identifier mutant} This technique eliminates some agent identity verifications from the protocol. 
Indeed, each communicating partner  sends identification information within each message. 
At reception, each partner verifies the correctness of the provided identity. 
If it is not correct the protocol closes the communication. 
Let us consider these reception instructions in HLPSL: $RCV(\{A,B\}_k)$ and $RCV(\{A,B'\}_k)$. 
In the second instruction we remark that B is primed which means that the value is new. 
Hence at reception this value cannot be checked against a  previously known name. 
So, any value including intruder provided one will be accepted. This can be exploited to mount an attack. 
In order to perform this mutation at the source code level, it suffices to  delete  identity verification. 

\paragraph{Nonce mutant}In order to verify the session freshness and the agent presence in a cryptographic protocol nonces 
are  random messages  generated by the communication partners. An agent creates its own nonces, sends them to the other parties and waits to get them back in some response. 
Let us consider these HLPSL instructions: $RCV(\{msg,Na\}_k)$ and $RCV(\{msg,Na'\}_k)$. In the first instruction the agent 
will compare the received nonce $Na$ with a previous copy that it possesses. 
In the second instruction, there is no verification since $Na'$ is considered to be new. 
So, this mutation creates a  flaw. 
\paragraph{}Mutation techniques permit one to focus the search for implementation vulnerabilities  
and   generate relevant attack trace scenarios. 
In our work we have used automatic mutation operators that affects specific security properties. 

\paragraph{Attack trace} Using  mutation techniques and  the CL-Atse Protocol Analyser ~\cite{clatse} 
we can generate multiple attack traces each one describing an attack scenario against a specific security property. An attack trace is given by a set of instructions. 
In our case, we aim to test the existence of the renegotiation  vulnerability in SSL/TLS protocol server implementation; 
A vulnerability in the way SSL and TLS protocols allow renegotiation requests may allow an attacker to inject plaintext into an application protocol stream. This could result in a situation where the attacker may be able to issue commands to the server that appear to be coming from a legitimate source ~\url{www.kb.cert.org/vuls/id/120541}. 
The following attack trace model describes the renegotiation process:  
 
\begin{tabbing}
$I\rightarrow B : start$\\
$I\rightarrow B: (I.Ni.Sid.Pi1)$\\
$B\rightarrow I: (Nb.Sid.Pb)$\\
$I\rightarrow B : pair \{crypt (Kb, PMS), crypt(IntruderK,Finished)\} $\\
$B\rightarrow I: crypt(IntruderK,Finished1)$\\
$I\rightarrow B: crypt((I.Ni.Sid.Pi2), IntruderK)$\\

\end{tabbing}
\section{A testing platform architecture}

\paragraph{}As discussed above, the attack trace is rather abstract and in order to be able to detect real attacks that affects protocol implementations we need a platform that provides:
\begin{itemize}
	\item[i)] Messages format conversion from a formal level to the implementation level
	\item[ii)] Real communications with the system under test. 
\end {itemize}

These are the main aspects of our solution. We introduce the platform architecture by describing their components, their functionalities and their interactions.  
\begin{figure}[H]
	\centering
		\includegraphics[width=11cm]{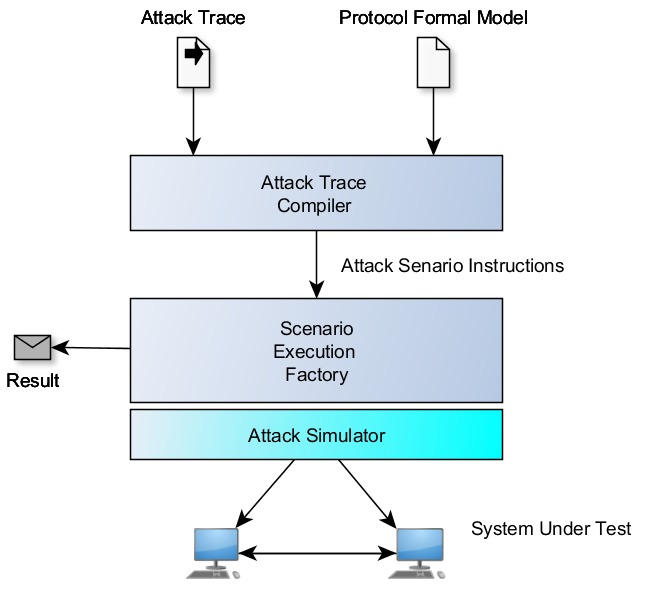}
	\caption{Platform Architecture}
	\label{fig:Platform}
\end{figure}
The platform architecture has three main components each with a specific role:
\\
\begin{itemize}
	\item[1.] Attack Trace Compiler:  identifies agents, messages types and operations. 
	\item[2.] Scenario Execution Engine: generates ( and verify) outgoing (resp. incoming) messages.
	\item[3.] Attack Simulator: simulates the scenario on real communication channels 
\end{itemize}

As shown in Figure~\ref{fig:Platform}, the platform takes as input the attack trace and 
the mutated model of the protocol under test, 
and  returns as a result an indication whether the considered attack on the examined implementation exists or not.

\subsection{Attack Trace Compiler:}
\begin{figure}[H]
	\centering
		\includegraphics[width=15cm]{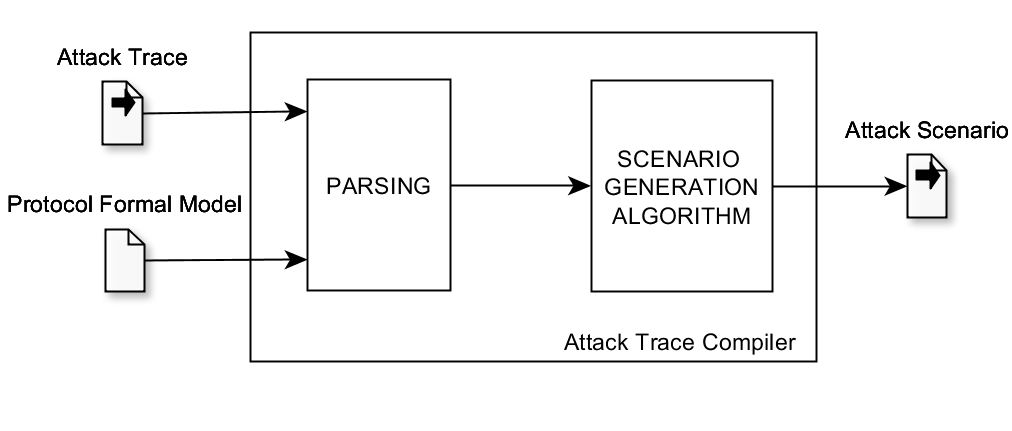}
	\caption{Attack Trace Compiler}
	\label{fig:Compiler}
\end{figure}
The Attack Traces Compiler  transforms an (abstract) attack trace into an (executable) attack scenario. 
This module provides an attack scenario which describes in detail the actions which should be done by the intruder while executing the attack.
The attack scenario is structured into steps and elementary instructions. Each step corresponds to an abstract attack trace elementary step.
This module takes as inputs an  attack trace and the protocol formal model written in HLPSL Language. 
The abstract trace  serves as a guide for the intruder implementation. 
The HLPSL model is used to collect some information such as the intruder initial knowledge.
The Attack Trace Compiler  identifies messages types and operations. 

The output is the attack trace scenario describing the intruder behaviour. It is a set of elementary steps used by the Scenario Execution Engine to implement this intruder. 
In order to explain well the process of compiling the attack trace, we give here a simple example of an attack trace treatment:
\begin{tabbing}
$Attack \ trace \ model :$\\
$i\rightarrow a: start$\\
$a\rightarrow i: crypt(kb,pair(Na,a))$\\
$i\rightarrow a: crypt(ka,pair(Na,pair(Nb,b)))$\\
$a\rightarrow i: crypt(kb,Nb) $\\
\end{tabbing}
\begin{tabbing} 
$Attack$\= $trace$\= $Scenario\ of\ I$\\
\>$Step\ -1:$\\
\>\> $0 = start = iknown$\\
\>\> $1 = a = iknown$\\
\>\> $2 = b = iknown$\\
\>\> $3 = ka = iknown$\\
\>\> $4 = kb = iknown$\\
\>\> $5 = ki = iknown$\\
\>\> $6 = inv(ki) = iknown$\\
\>\> $7 = i = iknown$\\
\>$Step\ 0:$\\
\>\> $!0 = start $\\
\>$Step\ 1:$\\
\>\> $?8= crypt(kb,pair(Na,a)) $\\
\>$Step\ 2:$\\
\>\> $!11= crypt(ka,pair(Na,pair(Nb,b))) $\\
\>\> $8="received\ at\ step:1"$\\
\>$Step\ 3:$\\
\>\> $?16= crypt(kb,Nb)$\\
\>\> $13="generated\ nonce\ at\ step:2"$\\
\>\> $15="generated\ nonce\ at\ step:2"$\\
\>\> $14=pair(15,2)$\\
\>\> $16=crypt(4,15)$\\
\>\> $12=pair(13,14)$\\
\>\> $11=crypt(3,12)$\\
\>$Step\ 4:$\\
\>\> $finish()$
\end{tabbing}
The steps are indexed. The first one  stores the  Initial Knowledge   (keys or agent's identities) deduced from the protocol model. We note that "?" (and "!") symbol refers to a receive (resp. send) operation.
The result of each operation is saved in an intermediate indexed variable (Xi). This will help the scenario execution engine 
to handle  the generated flow of information.

\paragraph{Scenario Generation Algorithm}

In order to generate an \emph{executable} attack scenario, for each step in $A \rightarrow B: M$ in the attack trace, where 
$A$ is an agent controlled by the intruder, one needs to check whether the intruder can compose $M$ from 
the pieces of knowledge he has at this step. 
Therefore we need to use a procedure to verify whether a message $t$ can be composed from a set of messages $E$ by the intruder simulator. The method provides A sequence of operations that allows one to construct a  given message.
When several sequences are possible, the choice does not matter for the remaining of our process.
The method is complete in the sense that if the message can be constructed
we derive (at least) one sequence of operations for that. 

This algorithm is well-known in protocol analysis (e.g.~\cite{np}) and it is 
 implemented in  Avantssar Platform \cite{avantssar} as part of the  Orchestration module. We have reused 
an optimized version  for our purpose.
The algorithm  provides us with the sequence of operations to apply in order to generate $t$ from $E$ whenever it is possible. 
This sequence of operations is represented as a term  $\overline{t}$. 
The proposed method  constructs $t$ from knowledge set $E$.
This is done by checking  which subterm in $E \cup \{t\}$ can be constructed. 
Several passes are necessary since after constructing some keys one can 
decrypt  more messages. Hence, iteratively, more and more messages get constructible.  
At the end of the main loop one checks whether the message  $t$ is in the set of constructible messages $K_1$.

\subsection{Scenario Execution Engine}
This module is responsible of translating the attack scenario from formal level to the implementation level. It ensures the association between abstract messages and concrete ones, stored in the Data Store module. Operation execution is held with the functionality provided by the Primitive Holder.
\paragraph{Primitive Holder} Our execution environment works at the implementation level. The exchanged messages are real network messages. Therefore, we have to relate abstract messages  with actual messages and operations. This is the main role of the Scenario Execution Handler.
 We can classify the attack scenario instructions into three categories:
\begin{itemize}
\item[1.] Message construction,
\item[2.] Message sending
\item[3.] Message receiving. 
\end{itemize}
To do this, we use cryptographic primitives ($crypt,\ pair$ and $unpair$) and network primitives ($send$ and $receive$). 
In fact, we define all the needed cryptographic operations in the Primitive Holder Module. Corresponding to the specification of the protocol SSL/TLS, this entity provides operations such as encryption, decryption, nonce generation, signature and concatenation. 
In order  to execute the whole scenario without  errors  the Primitive Holder provides  all the possible operations accepted by the protocol implementation.    
\paragraph{Data Store} Message creation depends from the scenario's previous steps (we consider stateful protocols). 
Hence all the messages handled by the platform are saved in the Data Store  in their real format and indexed way. This  facilitates  data processing. 
The Scenario Execution Handler, retrieve data by providing their index. The same technique is used while storing the information. 
The Data Store contains also all objects required for intermediate computations 
like  encryption keys, data nonces, agent identities and  submessages. 
\paragraph{Scenario Execution Handler} This module handles the instantiation of abstract operations by concrete executable ones. Taking as input,
the elementary steps of an attack scenario, it processes  each instruction in order to identify the operation to perform 
and its arguments. It interacts with the Primitive Holder module to execute cryptographic operations and with the Data store module to save or retrieve operation arguments. 
Here we give the algorithm that describes  the interactions of the different modules:\\\

\begin{algorithm}[H]
   \caption{Scenario Execution Handler}
   \label{alg:ScExprocess}
   \KwIn{Instruction}
   \KwOut{Request\ to\ another\ component}
   Let  I\ contain\ instruction\ value\;

  \textbf{Case} \{I\ is\ send(Xi)\}  \textbf{then} \\ \hspace{1 cm} Get\ data\ from\ the\ Data\ Store\ at\ position\ i ;\\
     	\hspace{1 cm} Call\ A-Simulator\ to\ send\ message \newline
 \textbf{Case} \{I\  is\ Xi=receive()\} \textbf{then} \\\hspace{1 cm} Call\ A$-$Simulator\ to\ get\ the\ received\ message ; \\
   		\hspace{1 cm}    Store\ the\ message\ on\ the\ Data\ Store\ at\ position\ i \newline
  \textbf{Case}\{I\ is\ Xi=operator(Xy,Xz)\} \textbf{then}\\ \hspace{1 cm} Get\ data\ from\ Data\ Store\ at\ positions\ y\ and\ z;\\
    		\hspace{1 cm}	Call\ the\ Primitive\ Holder\ to\ execute\ the\ primitive;\\
    		\hspace{1 cm}	Store\ the\ message\ on\ the\ Data\ Store\ at\ position\ i\
      		\newline
   \textbf{Case}\{I\ is\ finish()\} \textbf{then}\\ \hspace{1 cm} Exit\ with\ success
\end{algorithm}

The last case of Algorithm \ref{alg:ScExprocess} is the message construction or decomposition. In both cases the Handler invokes the Data Store and the Primitive Holder modules 
 Consider for instance the instruction $X1 = Crypt(X2, X3)$.  First, the Scenario Execution Handler collects the arguments by requesting them from the Data Store through the data buffer. Then, it requests the Primitive Holder to construct the message. Finally, the latter's response is stored at the result position  $X1$ in the Data Store.
At the end of the attack scenario execution, the Scenario Execution Engine sends a ``finish'' message. This means that the attack works on the tested implementation. 
\subsection{Attack Simulator}
After transforming a message from the formal to the real format, the Scenario Execution Handler processes the sending and the receiving operations. In such case, it sends a request to the Attack Simulator module which is the interface of the platform with the external environment.
 At the formal level, the intruder interacts with the other agents via channels.  Moreover, he has the capacity either to intercept messages 
in the passive attack case or to block and send (altered) messages when he wants to impersonate other agents. This is the active attack case. 
In both cases, Attack Simulator should provide the following functions:
\begin{itemize}
\item Create the real communication channels.
\item Send messages.
\item Intercept messages.
\item Block messages.
\item Redirect messages.
\item Create agent (depending on the attack scenario)
\end{itemize}
This module represents our platform interface with the system under test. It's responsible not only for ensuring real communication but also 
for validating the execution of the whole attack scenario.
\paragraph{Architecture and Functioning} The execution environment is composed of the agents and the intruder. They interact over the network through the communication channels.The agents are protocol process running at a separate hosts.  We propose to simulate attack scenarios in a local network to get a full control on different instantiated agents and also to avoid any low violation caused if the attack scenario succeed.
The attack simulator, creates agents depending on the attack scenario.This is done manually . For instance, in the case of a Man In The middle attack, there is three agents A,B and I. A is a legitimate client, B is a legitimate server and I simulates our intruder.
If we assume that B is the system under test. Therefore, the attack simulator will create two agents; a legitimate one A and malicious one I.  
\paragraph{Attack validation} Attack validation is the most important step of the test. 
The simulator logs all the exchanged traffic between the platform and the System Under Test. 
Indeed, every transiting packet is checked especially the received ones. They may be either a response to a previous request from the platform or an error message. The error messages are provided by the protocol owner. We put them in a configuration file. If the protocol response is one of them, we are sure that the protocol implementation does not accept the attack scenario. The formal attack does not exist in implementation. 
Otherwise, if the packet source is the Scenario Execution Engine and its content is a ``finish'' message. 
Therefore, our attack scenario execution confirms a  security flaw on  the implementation. 
\section{Conclusion}

We have introduced a  platform architecture for protocol blackbox testing. The platform exploits an attack trace to 
guide the generation of an  intruder implementation. 
Then, using a simulation module, the attack scenario can be played on a real protocol 
implementation and  allow us to know if this implementation is vulnerable. 
\bibliographystyle{plain}
\providecommand{\urlalt}[2]{\href{#1}{#2}} \providecommand{\doi}[1]{doi:\urlalt{http://dx.doi.org/#1}{#1}}

\end{document}